\documentstyle[preprint,aps,epsfig]{revtex}
\tightenlines
\begin{document}
\draft
\preprint{\begin{minipage}[b]{1.5in}
          UK/TP 99-17\\
          IITAP-99-015\\
          hep-ph/9911322
          \end{minipage}}
\vspace{0.2in}

\title{{\bf $x_F$} dependence of the Drell-Yan transverse momentum
broadening}
\author{Xiaofeng Guo$^1$\footnote{
Email: gxf@ruffian.pa.uky.edu}, Xiaofei Zhang$^{2}$, and Wei Zhu$^{3,4}$}
\address{$^1$Department of Physics and Astronomy,
             University of Kentucky,\\
             Lexington, Kentucky 40506, USA\\
         $^2$Department of Physics and Astronomy,
             Iowa State University \\
             Ames, Iowa 50011, USA \\
         $^3$Department of Physics,
             East China Normal University\\
             Shanghai 200062, China \\
         $^4$International Institute of Theoretical and Applied
             Physics, Iowa State University\\
             Ames, Iowa 50011, USA}
\date{January 8, 2000}
\maketitle
\begin{abstract}
We analyze $x_F$ dependence of the Drell-Yan transverse momentum
broadening in hadron-nucleus collisions. In terms of generalized
factorization theorem, we show that the $x_F$ dependence of 
the transverse momentum broadening, 
$ \Delta \langle q_T^2 \rangle (x_F) $, can be
calculated in perturbative QCD. We 
demonstrate that $\Delta \langle q_T^2 \rangle (x_F)$ is a good
observable for studying the effects of initial-state multiple
scattering and extracting quark-gluon correlation functions.
\end{abstract}
\vspace{0.2in}

Parton multiple scattering is responsible for many interesting and
important phenomena in high energy collisions involving nucleus,
such as transverse momentum broadening, energy loss, as well as
the nuclear suppression of quarkonium states. Understanding
the mechanism of parton multiple scattering and its effects is
crucial for making precise predictions of the nuclear suppression
of quarkonium productions, which maybe a potential signal for the
quark-gluon plasma in relativistic heavy ion collisions \cite{Satz}.
Parton multiple scattering can happen both at the initial state or 
at the final state. The Drell-Yan pair production in hadron-nucleus
collisions provide an excellent place to study effects of
initial state parton multiple scattering. The nuclear dependence
in the Drell-Yan transverse momentum spectrum for the large $q_T$
region has been studied in QCD perturbation theory
\cite{Guo1,Stein}. In terms of generalized factorization theorem
in QCD \cite{QS_fac}, the effects of multiple scattering can be
expressed in terms of multiparton correlation functions
\cite{LQS1}, which are as fundamental as the parton distributions.
However, these parton correlation  functions are not well
determined yet.

In this letter, we derive the $x_F$ dependence of the Drell-Yan
transverse momentum broadening. We show that $\Delta \langle q_T^2
\rangle (x_F)$ can be used as a good observable to study the
effects of initial-state multiple scattering and parton energy
loss. In addition, it can be used as an excellent observable for
extracting information on multi-parton correlation functions.

Consider the Drell-Yan process in hadron-nucleus collisions,
$  h(p') + A(p) \rightarrow \ell^+\ell^-(q) + X $, where
$q$ is the four-momentum for the virtual photon $\gamma^*$
which decays into the lepton pair.  $p'$ is the momentum for
the incoming beam hadron and $p$ is the momentum per nucleon
for the nucleus with the atomic number $A$.
Let $q_T$ be the transverse momentum of the Drell-Yan pair, we
define the averaged transverse momentum square as
\begin{equation}
\langle q_T^2\rangle ^{hA}=
\left. \int dq_T^2 \cdot q_T^2 \cdot
\frac{d\sigma_{hA}}{dQ^2dq_T^2} \, \right/
\frac{d\sigma_{hA}}{dQ^2} \ .
\label{qt2}
\end{equation}
In Eq.~(\ref{qt2}), $Q$ is the total invariant mass of the lepton
pair with $Q^2=q^2$.  Since single hard scattering is localized in
space, only multiple scattering (at least, double scattering) are
sensitive to  the nuclear size (or $A^{1/3}$ type
dependence).  Therefore, in order to extract the effect due to multiple
scattering, we introduce the nuclear enhancement of the Drell-Yan
$\langle q_T^2 \rangle$ as
\begin{equation}
\Delta \langle q_T^2\rangle
\equiv \langle q_T^2 \rangle ^{hA}
      -\langle q_T^2 \rangle ^{hN} \ ,
\label{dydqt2}
\end{equation}
which is often called the transverse momentum broadening. The
broadening 
of the transverse momentum square defined in Eq.~(\ref{dydqt2}) should
be sensitive to parton multiple scattering between nucleons inside
a large nucleus.

In a perturbatively calculable hard scattering process, having an
extra scattering between physical partons is
suppressed by a power of the hard scale \cite{LQS1}. Therefore,
multiple scattering in momentum space between physical 
partons correspond to an expansion in power series of $1/Q^2$.  In
this letter, we limit ourselves to double scattering in momentum
space. With only single and double scattering, the broadening of the 
Drell-Yan transverse momentum square, $\Delta\langle
q_T^2\rangle$, can be parameterized as 
\begin{equation}
\Delta\langle q_T^2\rangle
= a + b\, A^{1/3}\ ,
\label{BA13}
\end{equation}
which is consistent to existing data \cite{E683,E772}.
In Eq.~(\ref{BA13}), $bA^{1/3}$-term represents
the contribution directly from the double scattering which is
explicitly proportional to the nuclear size ($\propto A^{1/3}$), 
with A the atomic weight of the nucleus target.

In Ref.~\cite{QS_fac}, Qiu and Sterman argued that the
factorization theorem for hadron-hadron scattering \cite{CSS_fac}
should also be valid at the first non-leading power in momentum
transfer, which is essential for systematically 
calculating the double scattering in QCD perturbation theory. 
According to this generalized factorization theorem, we can expand 
the numerator in Eq.~(\ref{qt2}) as
\begin{eqnarray}
\int d q_{T}^2 \, q_{T}^2 \,
\frac{d\sigma_{hA}}{ dQ^2 d q_{T}^2}
&=&
\sum_{a,b} \phi_{a/A}(x)\otimes
   C^{(0)}_{ab\rightarrow l\bar{l}}\left(x,Q^2\right)
           \otimes  \phi_{b/h}(x')
\nonumber \\
&+&\frac{1}{Q^2}
\sum_{a,b} \left[
           T_{a/A}(x) \otimes
   C^{(2)}\left(x,Q^2\right)
           \otimes \phi_{b/h}(x') \right.
\nonumber \\
&& {\hskip 0.4in} \left.
          +\phi_{a/A}(x) \otimes
\bar{C}^{(2)} \left(x,Q^2\right)
           \otimes T_{b/h}(x') \right]
\, +\, ... 
\nonumber \\ 
&\equiv & H^{(0)}_A + H^{(2)}_A + \bar{H}^{(2)}_A + ...\ .
\label{Hexp}
\end{eqnarray}
where $\otimes$ represents convolutions over partonic momenta, and
``...'' represents the terms that are suppressed by higher powers
of $1/Q^2$. In Eq.~(\ref{Hexp}), $C^{(0)}$, $ C^{(2)}$, and
$\bar{C}^{(2)}$ are perturbatively calculable hard parts.
$\phi_{b/h}(x')$ is the parton distribution of the beam hadron, and
$\phi_{a/A}(x)$ is the parton distribution in the nucleus
normalized by the atomic number A. $T_{a/A}(x)$ and $T_{b/h}(x')$
are the four-parton correlation functions \cite{LQS1} in the
nucleus and the beam hadron, respectively.

Because of the well-known EMC effect, as well as effects of nuclear
shadowing and Fermi motion, the nuclear dependence of the $\phi_{a/A}$
is nontrivial.  However, if we parameterize the effective nuclear
parton distribution $\phi_{a/A}$ into $A^{\alpha}$ times
corresponding nucleon parton distributions, we found a very small 
power of $\alpha$ for a wide range of $x$ \cite{GQ-qt}.  Taking the
EKS98 parameterization of nuclear parton distributions \cite{Eskola}
as an example, we found that $\alpha \sim \pm (0.02-0.03)$ for the
$x$-range relevant to this study, 
which is much smaller than $1/3$ for the $A^{1/3}$-type
enhancements.  Therefore, $H_A^{(0)}$ in Eq.~(\ref{Hexp}), which is
proportional to $\phi_{a/A}$, should have a very weak nuclear
dependence. Similarly, $\bar{H}_A^{(2)}$ in Eq.~(\ref{Hexp}) also
has a weak nuclear dependence.  On the other hand, the nuclear
parton correlation function $T_{a/A}$ has an explicit dependence
on the nuclear size ($\propto A^{1/3}$)\cite{LQS1,LQS2}, so as the
$H_A^{(2)}$ in Eq.~(\ref{Hexp}).

According to the factorization theorem \cite{CSS_fac}, the denominator
in Eq.~(\ref{qt2}) can also be expand in terms of power series:
\begin{eqnarray}
\frac{d\sigma_{hA}}{dQ^2} &=& \sum_{a,b} \phi_{a/A}(x,\mu^2)\,
\otimes\, \frac{d\hat{\sigma}^{(0)}_{ab\rightarrow
l\bar{l}}}{dQ^2}
       \left(x,x',\mu^2/Q^2,\alpha_s(\mu^2)\right) \,
\otimes\, \phi_{b/h}(x',\mu^2) \left[ 1+
O\left(\frac{1}{Q^2}\right)\right]\ \nonumber
\\ &\equiv & \sigma_A^{(0)}\,\left[ 1+
O\left(\frac{1}{Q^2}\right)\right]\ , 
\label{D0}
\end{eqnarray}
where $\mu$ represents both renormalization and factorization scale.
In Eqs.~(\ref{Hexp}) and (\ref{D0}), all quantities are normalized
by the atomic number $A$.
Substituting Eqs.~(\ref{Hexp}) and (\ref{D0}) into Eq.~(\ref{qt2}), 
we obtain 
\begin{equation}
\langle q_{T}^2\rangle^{hA} \approx
\frac{H^{(0)}_A+H^{(2)}_A}{\sigma^{(0)}_A}
\,\left[ 1+ O\left(\frac{A^0}{Q^2}\right)\right] \ .
\label{expLTA}
\end{equation}
In deriving Eq.~(\ref{expLTA}), we kept only terms up to
$O(A^{1/3}/Q^2)$, and dropped all power correction terms without
the $A^{1/3}$ nuclear enhancement.  Similarly, for a nucleon target,
we have
\begin{equation}
\langle q_{T}^2\rangle^{hN} \approx
\frac{H^{(0)}_N}{\sigma^{(0)}_N} \,
\left[ 1+ O\left(\frac{A^0}{Q^2}\right) \right]\ .
\label{expLTN}
\end{equation}
Substituting above Eqs.~(\ref{expLTA}) and (\ref{expLTN}) into our
definition of the nuclear broadening of the transverse momentum square
in Eq.~(\ref{dydqt2}), we derive
\begin{equation}
\Delta\langle q_T^2\rangle \approx
\left[\frac{H^{(0)}_A}{\sigma^{(0)}_A} -
      \frac{H^{(0)}_N}{\sigma^{(0)}_N} \right] +
\frac{H^{(2)}_A}{\sigma^{(0)}_A} \ ,
\label{DLT2com}
\end{equation}
where we neglected terms of $O(A^0/Q^2)$. The first term in
Eq.~(\ref{DLT2com}) should have a very weak nuclear dependence due
to the fact that effective nuclear parton distributions have a
very week $A^\alpha$ dependence.  Therefore, the first term
contributes to 
the $a$-term in Eq.~(\ref{BA13}). In addition, the first term
should be numerically very small due to the cancellation between
the two terms inside the bracket.  On the other hand, the second term in
Eq.~(\ref{DLT2com}) represents the double scattering contribution
and it gives the $A^{1/3}$-type enhancement, and therefore, it
contributes to the $bA^{1/3}$-term in Eq.~(\ref{BA13}).

At the leading order, the inclusive cross section is \cite{Fields}
\begin{equation}
\sigma^{(0)}_A \equiv
\frac{d\sigma_{hA \rightarrow \ell^+ \ell^-}}{dQ^2}
= \sigma_0 \, \sum_{q}\, e_q^2
\int\, dx'\, \phi_{\bar{q}/h}(x') \,
\int\, dx\, \phi_{q/A}(x) \, \delta (Q^2-xx's) \ ,
\label{dysingle}
\end{equation}
with $s=(p+p')^2$ and the Born cross section
\begin{equation}
\sigma_0=\frac{4\pi \alpha_{em}^2}{9Q^2} \ .
\label{sigma0}
\end{equation}
Also at the leading order, the Drell-Yan differential cross
section is given by
\begin{equation}
\frac{d\sigma_{hA \rightarrow \ell^+ \ell^-}}{dQ^2 dq_T^2}
=\frac{d\sigma_{hA \rightarrow \ell^+ \ell^-}}{dQ^2} \, \delta (q_T^2)
\ .
\label{LO-spectrum}
\end{equation}
Therefore, $H_A^{(0)}$ and $H_N^{(0)}$ in Eq.~(\ref{DLT2com})
vanish at the leading order, because $\int dq_T^2\, q_T^2\,
\delta(q_T^2)=0$, and consequently, the first term in
Eq.~(\ref{DLT2com}), or the $a$-term in Eq.~(\ref{BA13}), vanishes at
the leading order in $\alpha_s$. 

The double scattering contribution to the Drell-Yan transverse
momentum broadening, $H_A^{(2)}$ in Eq.~(\ref{DLT2com}),  was
derived in Ref.~\cite{Guo2}. At the leading order in $\alpha_s$, it is
given by \cite{Guo2}
\begin{eqnarray}
 H_A^{(2)}
&=& \sigma_0  \left(\frac{4\pi^2 \alpha_s}{3} \right) \sum_{q}
\int\, dx'\, \phi_{\bar{q}/h}(x') \int\, dx\, T_{q/A}^{(I)}(x)\,
\delta (Q^2-xx's) \ , \label{qt2Db}
\end{eqnarray}
where the four-parton correlation function is defined as
\begin{eqnarray}
T_{q/A}^{(I)}(x) &=&
 \int \frac{dy^{-}}{2\pi}\, e^{ixp^{+}y^{-}}
 \int \frac{dy_1^{-}dy_{2}^{-}}{2\pi} \,
      \theta(y^{-}-y_{1}^{-})\,\theta(-y_{2}^{-}) \nonumber \\
&\ & \times \,
     \frac{1}{2}\,
     \langle p_{A}|F_{\alpha}^{\ +}(y_{2}^{-})\bar{\psi}_{q}(0)
                  \gamma^{+}\psi_{q}(y^{-})F^{+\alpha}(y_{1}^{-})
     |p_{A} \rangle \ .
\label{dyTq}
\end{eqnarray}
These parton correlation functions are not well measured yet. By
comparing the operator definitions of the correlation functions and
the definitions of the normal twist-2 parton distributions, Luo,
Qiu, and Sterman (LQS) proposed the following model
\cite{LQS1,LQS2}:
\begin{equation}
T_{f/A}(x)=\lambda^2 A^{1/3} \phi_{f/A}(x) \ ,
\label{TiM}
\end{equation}
where $\lambda$ is a free parameter to be fixed by experimental
data, and was estimated in Ref.~\cite{Guo2} as
$\lambda^2=0.01$~GeV$^2$ by using the Drell-Yan data
from NA10 and E772 experiment \cite{E772,NA10}.

In order to obtain the leading order $x_F$-dependence of
$\Delta\langle q_T^2 \rangle$, we multiply $\delta(x_F-(x'-x))\,
dx_F$ to the right-hand side of Eq.~(\ref{qt2Db}), and obtain
\begin{equation}
\frac{dH_A^{(2)}(x_F)}{dx_F} = \sigma_0  \left(\frac{4\pi^2
\alpha_s}{3} \right) \sum_{q} \phi_{\bar{q}/h}(x_1) \,
T_{q/A}^{(I)}(x_2)\, \frac{1}{(x_1+x_2)s}\, ,
\label{H2xf}
\end{equation}
with 
\begin{equation}
x_1=(\sqrt{x_F^2+4\tau}+x_F)/2\, , 
\quad \mbox{and}\quad
x_2=(\sqrt{x_F^2+4\tau}-x_F)/2\, ,
\label{x1x2}
\end{equation} 
where $\tau=Q^2/s$. 
Combining Eqs.~(\ref{DLT2com}), (\ref{dysingle}), and (\ref{H2xf}),
we obtain the $x_F$ dependence of the Drell-Yan transverse momentum
broadening $\Delta \langle q_T^2
\rangle$ at the leading order in $\alpha_s$:
\begin{eqnarray}
\frac{d\Delta \langle q_T^2 \rangle (x_F)}{d x_F}
&=&\left(\frac{4\pi^2 \alpha_s}{3} \right)\cdot \frac{\sum_{q} \,
e_q^2  \, \phi_{\bar{q}/h}(x_1)\, T^{(I)}_{q/A}(x_2) /(x_1+x_2)}
{\sum_{q}\, e_q^2 \int dx' \, \phi_{\bar{q}/h}(x')\,
\phi_{q/A}(\tau /x) /x'}  \ .
\label{dyqt2b}
\end{eqnarray}
From Eq.~(\ref{dyqt2b}), we see that $d\Delta \langle q_T^2
\rangle/d x_F$ directly depends on the quark-gluon correlation
functions. Therefore, the broadening is a very good observable for
measuring quark-gluon correlation functions.

If we use LQS model for $T_{q/A}$ given in Eq.~(\ref{TiM}), from
Eq.~(\ref{dyqt2b}), we can derive a much simpler expression for the
broadening,
\begin{equation}
\frac{d\Delta \langle q_T^2 \rangle}{d x_F} =\left( \frac{4\pi^2
\alpha_s}{3} \right)\, \lambda^2 A^{1/3}\, \left.
\frac{d\sigma}{dQ^2d x_F} \right/ \frac{d\sigma}{dQ^2} \ .
\label{dyqtxf}
\end{equation}
From Eq.~(\ref{dyqtxf}), we can see that the Drell-Yan transverse
momentum broadening should have similar $x_F$ dependence  as the
differential cross section $d\sigma /dQ^2d x_F$ if the LQS model
for $T^{(I)}_{q/A}(x)$ is valid. Therefore, by comparing the $x_F$
dependence of $d\Delta \langle q_T^2 \rangle/d x_F$ and $d\sigma
/dQ^2d x_F$, we can provide an immediate test of LQS model for the 
correlation functions.  We emphasize that even if LQS model is
not a good approximation for the quark-gluon correlation
functions, measuring the
$x_F$-dependence of the nuclear broadening provides excellent
information for extracting the quark-gluon correlation functions
$T_{q/A}(x)$ directly, as shown in Eq.~(\ref{dyqt2b}).

In the following, we use Eq.~(\ref{dyqtxf}) to obtain the
numerical estimates of the $x_F$ dependence of the Drell-Yan
transverse momentum broadening. Although the  value of $\lambda^2$
for the correlation function is not well determined, a different
value of $\lambda^2$ corresponds to a simple adjustment to the
overall normalization. Therefore, the uncertainty in the value of
$\lambda^2$ should not affect our following discussions.

In obtaining our following numerical results, we use the CTEQ4L
distribution as the quark distributions in the nucleon. For effective
quark distributions in the nucleus, we define
$q_{i/A}(x)=q_{i/p}(x)R_i(x,A)$, and use EKS98 for the
parameterizations of $R_i(x,A)$ \cite{Eskola}, which fit the data
well.  We choose the renormalization and factorization scale to be
$\mu=Q$, and choose the incoming beam energy $p=800$~GeV which is 
the energy used by the Fermilab experiments 
\cite{E772,E-loss,JPsi,Moss}.

\begin{figure}
\begin{minipage}[t]{2.0in}
\epsfig{figure=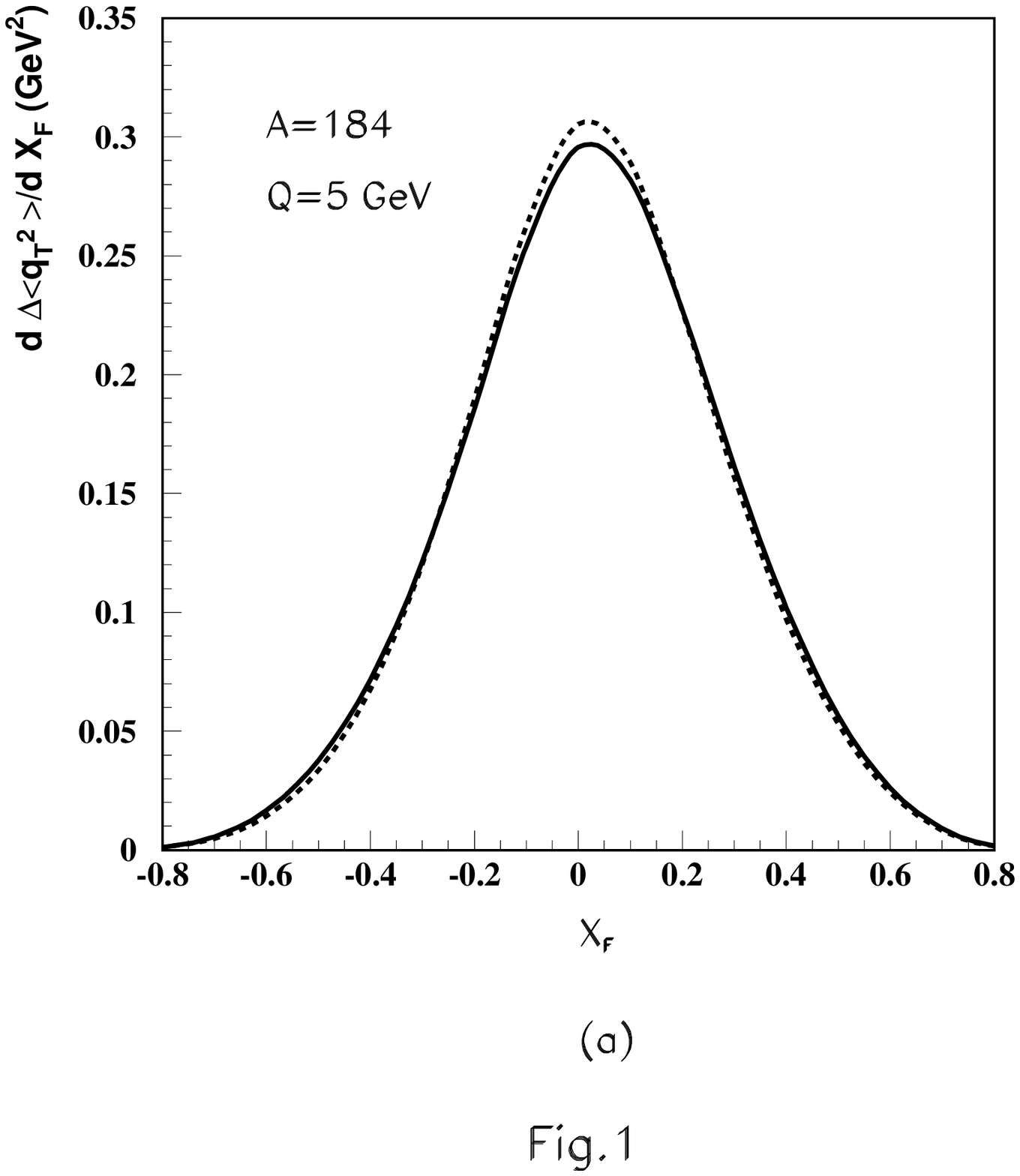,width=1.0in}
\end{minipage}
\hfill
\begin{minipage}[t]{2.0in}
\epsfig{figure=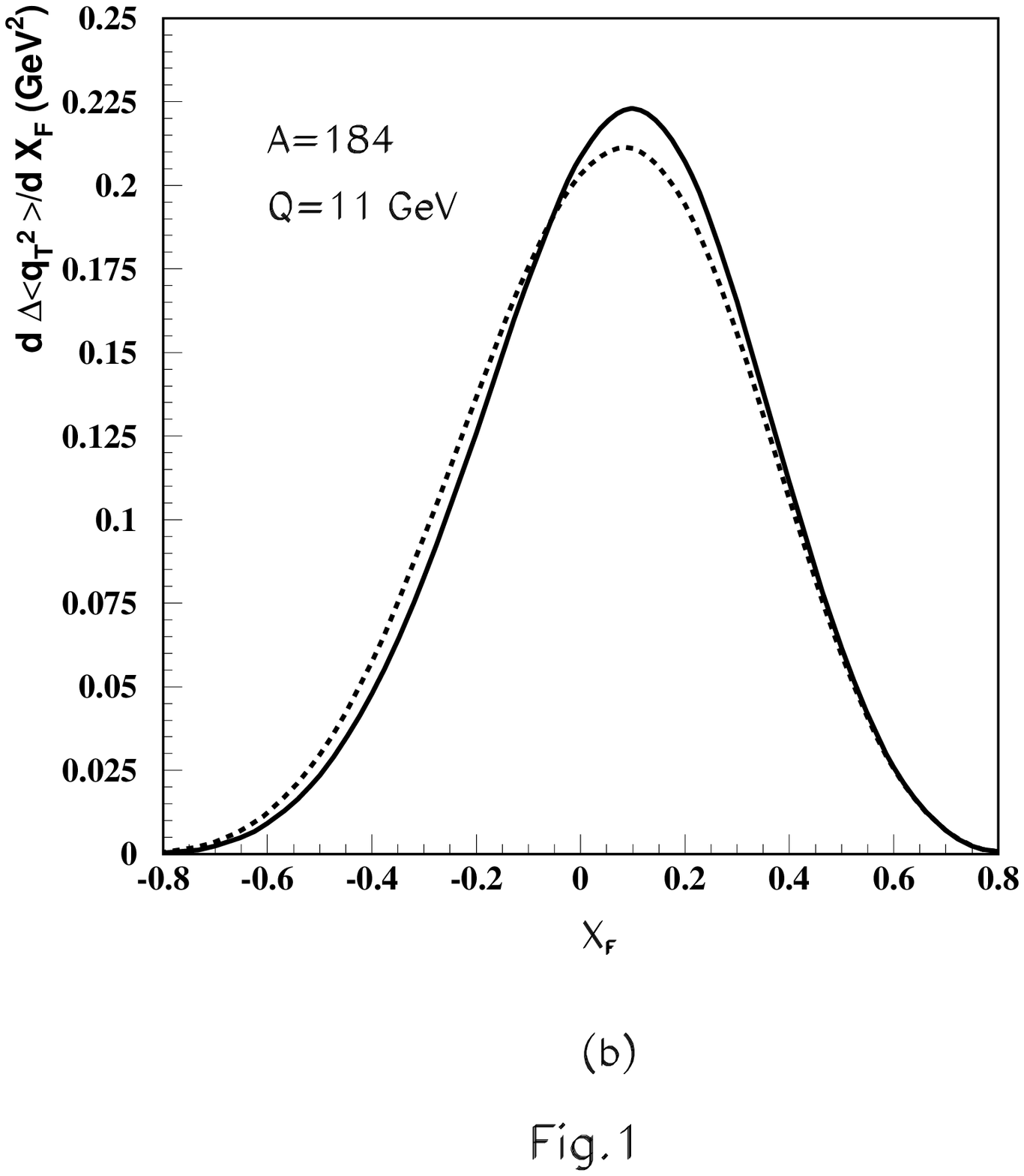,width=1.0in}
\end{minipage}
\caption{The transverse momentum broadening of the Drell-Yan pair, 
$d \Delta \langle q_T^2 \rangle / dx_F$, 
as a function of $x_F$ for $800$ GeV proton beam on nuclear target 
A=184, at $Q=5$ GeV (a) and $Q=11$ GeV (b).} 
\label{fig1}
\end{figure}

In Fig.~\ref{fig1}, we plotted $d\Delta \langle q_T^2 \rangle/d
x_F$ in Eq.~(\ref{dyqtxf}) as a function of $x_F$ with $A=184$ and
$Q=5$~GeV and 11~GeV, respectively. The dotted lines correspond to
EKS98 parameterizations of effective nuclear parton distributions.  In 
order to separate the nuclear dependence caused by multiple scattering
and that caused by the effective nuclear parton distributions, 
we also plotted $d\Delta \langle q_T^2 \rangle /d x_F$ in solid
curves with $R_i(x,A)=1$.  The difference between the solid and dotted
lines is a direct consequence of the difference between nucleon and
nuclear parton distributions.  As shown in Eq.~(\ref{dyqt2b}), $x_2$
represents the momentum fraction of a quark (or antiquark) from the
nuclear target.  At $x_F=0$, we have $x_2=Q/\sqrt{s} \approx 0.13$ for
$Q=5$~GeV, and $\approx 0.28$ for $Q=11$~GeV.  It is clear that
$Q=5$~GeV and $Q=11$~GeV cover very different range of $x_2$.  For
$Q=5$~GeV, $x_F>0$ corresponds to $x_2<0.1$ or corresponds to the
shadowing region, while $x_F<0$ covers the region of EMC effect. 
On the other hand, the Drell-Yan pairs of $Q=11$~GeV are not sensitive to
the nuclear shadowing at all.  The entire range of $x_F$ values matches
the range of EMC effect which include the antishadowing region ($0.1
\leq x_2 \leq 0.2$), and the EMC suppression region ($0.2\leq x_2 \leq
0.7$), as well as the Fermi motion region for larger $x_2$.  
Such dependence in effective nuclear quark distributions are
clearly shown in Fig.~\ref{fig1}.  In Fig.~\ref{fig1}a, the dotted
line is above the solid line in the central region of $x_F$ due to the
fact that $x_2$ is in the antishadowing region.  When $x_F$ is
positive, the dotted line is below the solid line because
$q_{i/A}(x_2)$ is now in the shadowing region.  When $x_F$ is less
than zero, the dotted line is again below the solid line due to the
fact that $x_2$ is now in the region of EMC suppression.
On the other hand, Fig.~\ref{fig1}b shows slightly different relation
between the dotted and solid line due to the fact that at $Q=11$~GeV,
$x_2$ covers a different range as shown in Fig.~\ref{fig1}.
Although the nuclear dependence in effective nuclear parton
distributions change the $x_F$ dependence of the Drell-Yan transverse
momentum broadening, the change (e.g., the difference between the
dotted and solid lines in Fig.~\ref{fig1}) is extremely small. 
Therefore, the $x_F$ dependence of the nuclear broadening, 
$\Delta \langle q_T^2 \rangle (x_F)$ is not very sensitive to nuclear
shadowing, and can be an excellent probe of parton multiple
scattering. 

In addition to the transverse momentum broadening, parton multiple 
scattering also causes the energy loss.  In recent experiments, 
the ratio of the cross section per nucleon from the Drell-Yan pair
production in $p-A$ collisions has been used as a direct
measurement of  the parton energy loss in nuclear medium
\cite{E-loss}.  However, as pointed out in Ref.~\cite{E-loss}, a
substantial fraction of the variation in the cross section ratios
versus $x_1$ comes from the shadowing of $\phi_{f/A}(x_2)$ at small
$x_2$, and therefore it is difficult to extract precise information 
on parton energy loss from the cross section ratios.
On the other hand, BDMPS showed that for a jet produced in nuclear 
matter, the energy loss per unit length
$-\frac{dE}{dz}$ and the transverse momentum broadening
$\Delta p_{\perp}^2$ has the following relation~\cite{BDMPS}: 
\begin{equation}
-\frac{dE}{dz}= \frac{\alpha_s N_c}{4} \, \Delta p_{\perp}^2 \ .
\label{loss-dpt}
\end{equation}
Similar to Eq.~(\ref{loss-dpt}), one can use the transverse momentum 
broadening of the Drell-Yan pair $ \Delta \langle q_T^2 \rangle (x_F)$
to estimate the parton energy loss due to the initial state multiple
scattering.  As we demonstrated in Fig.~\ref{fig1}, the 
variation in the shape of $\Delta \langle q_T^2 \rangle (x_F)$ 
due to the shadowing is small. Therefore, 
$\Delta \langle q_T^2 \rangle (x_F)$ is also a good probe for the parton
energy loss.

Cross sections on nuclear targets are often parameterized as
$A^\alpha$ times corresponding cross sections on a nucleon target.
In principle, the power $\alpha$ can be a function of $A$, $q_T$,
$x_F$, as well as other physical observables.  The Nuclear dependence
of $\alpha$ is often used to study the effects
of multiple scattering. In recent experimental studies of the
J/$\Psi$ and $\Psi'$ suppression in $p-A$ collisions \cite{JPsi},
data was presented in terms of  $\alpha$ as a function of the 
transverse momentum $p_T$ in different $x_F$ regions.  It was
found \cite{JPsi} that the shapes of $\alpha(p_T)$ are 
very similar  in different regions of $x_F$, and it was
concluded \cite{JPsi} that the parton energy loss is independent of 
the $c\bar{c}$ energy.

Using our result of $\Delta \langle q_T^2 \rangle (x_F)$, we can
also estimate the $\alpha$ as a function of $q_T$ in different
$x_F$ region for the Drell-Yan production. We define the parameter
$\alpha(q_T)$ for the Drell-Yan production as
\begin{equation}
\frac{d\sigma_{hA}}{dQ^2 dq_T^2} =A^{\alpha(q_T)} \times
\frac{d\sigma_{hN}}{dQ^2 dq_T^2} \label{alpha}\ .
\end{equation}
If $q_T$ is not too large, the $q_T$ spectrum of the Drell-Yan pairs
can be approximately 
parameterized by a Gaussian form \cite{Ellis},
\begin{equation}
\frac{d\sigma_{hN}}{dQ^2 dq_T^2} \propto \frac{1}{\langle q_T^2
\rangle^{hN}} \, \exp\left[-\frac{q_T^2}{\langle q_T^2
\rangle^{hN}}\right] \ ;
\label{qtexpN}
\end{equation}
and
\begin{equation}
\frac{1}{A}\frac{d\sigma_{hA}}{dQ^2 dq_T^2} 
\propto \frac{1}{\langle q_T^2
\rangle^{hA}} \, \exp\left[-\frac{q_T^2}{\langle q_T^2
\rangle^{hA}}\right] \ .
\label{qtexpA}
\end{equation}
Substituting Eqs.~(\ref{qtexpN}) and (\ref{qtexpA}) into 
Eq.~(\ref{alpha}), we derive
\begin{equation}
\alpha(q_T) =1+\, 
\frac{1}{\ln(A)}\, \left[
\ln\left(\frac{1}{1+\chi}\right) 
+ \frac{\chi}{1+\chi}\, \frac{q_T^2}{\langle q_T^2 \rangle^{hN}}
\right]\ ,
\label{alpha-qt}
\end{equation}
where $\chi \equiv \Delta \langle q_T^2 \rangle / 
\langle q_T^2 \rangle^{hN}$.  In deriving Eq.~(\ref{alpha-qt}), we
used Eq.~(\ref{dydqt2}).  In order to estimate the value of
$\alpha(q_T)$, we use the value $\langle 
q_T^2 \rangle^{hN} =1.2~$GeV$^2$ for the cross section per nucleon
\cite{Moss}. And we use Eq.~(\ref{dyqtxf}) to integrate over
different ranges of $x_F$ value to obtain the $\Delta \langle q_T^2
\rangle$ in different ranges of $x_F$.  For $A=184$, we choose three
different $x_F$ ranges, which are the same as the regions used in 
Ref.~\cite{JPsi}, small $x_F$ (SXF: $-0.1 \leq x_F \leq 0.3$),  
intermediate $x_F$ (IXF: $0.2 \leq x_F \leq 0.6$), and 
large $x_F$ (LXF: $0.3 \leq x_F \leq 0.93$).  For these three $x_F$
regions, we obtain the corresponding
values of $\Delta \langle q_T^2 \rangle$ as 
[0.10, 0.043, 0.026]~GeV$^2$ for $Q=5$~GeV, and 
[0.078, 0.043, 0.027]~GeV$^2$ for $Q=11$~GeV, respectively.
It is clear that the value of $\chi$ should be very small, and the
power $\alpha(q_T)$ in Eq.~(\ref{alpha-qt}) can be approximated as 
\begin{equation}
\alpha(q_T) \approx 
1+\, \frac{\chi}{\ln(A)}\, \left[
-1 + \frac{q_T^2}{\langle q_T^2 \rangle^{hN}} \right]\, .
\label{alpha-qt0}
\end{equation}
From Eq.~(\ref{alpha-qt0}), we conclude that if $q_T$ is not too
large, $\alpha(q_T)$ should show linear dependence on $q_T^2$ (or 
quadratic dependence on $q_T$).  For the Drell-Yan
production, we expect an extremely small coefficient due to a weak
transverse momentum broadening (or a small $\chi$ value in
Eq.~(\ref{alpha-qt0})). 

In the above discussion, we used the same averaged value of 
$\langle q_T^2 \rangle^{hN}$ for different $x_F$ ranges. 
In principle $\langle q_T^2 \rangle^{hN}$ should also have 
$x_F$ dependence due to kinematics. The larger $\left| x_F \right|$, 
the smaller $\langle q_T^2 \rangle^{hN}$.  Hence 
$\langle q_T^2 \rangle^{hN}$ should have similar $x_F$ dependence 
as $\Delta \langle q_T^2 \rangle (x_F)$ shown in Fig.~\ref{fig1}. 
Therefore, $\chi=\Delta \langle q_T^2 \rangle / 
\langle q_T^2 \rangle^{hN}$ should have even smaller dependence on
$x_F$.  

\begin{figure}
\epsfig{figure=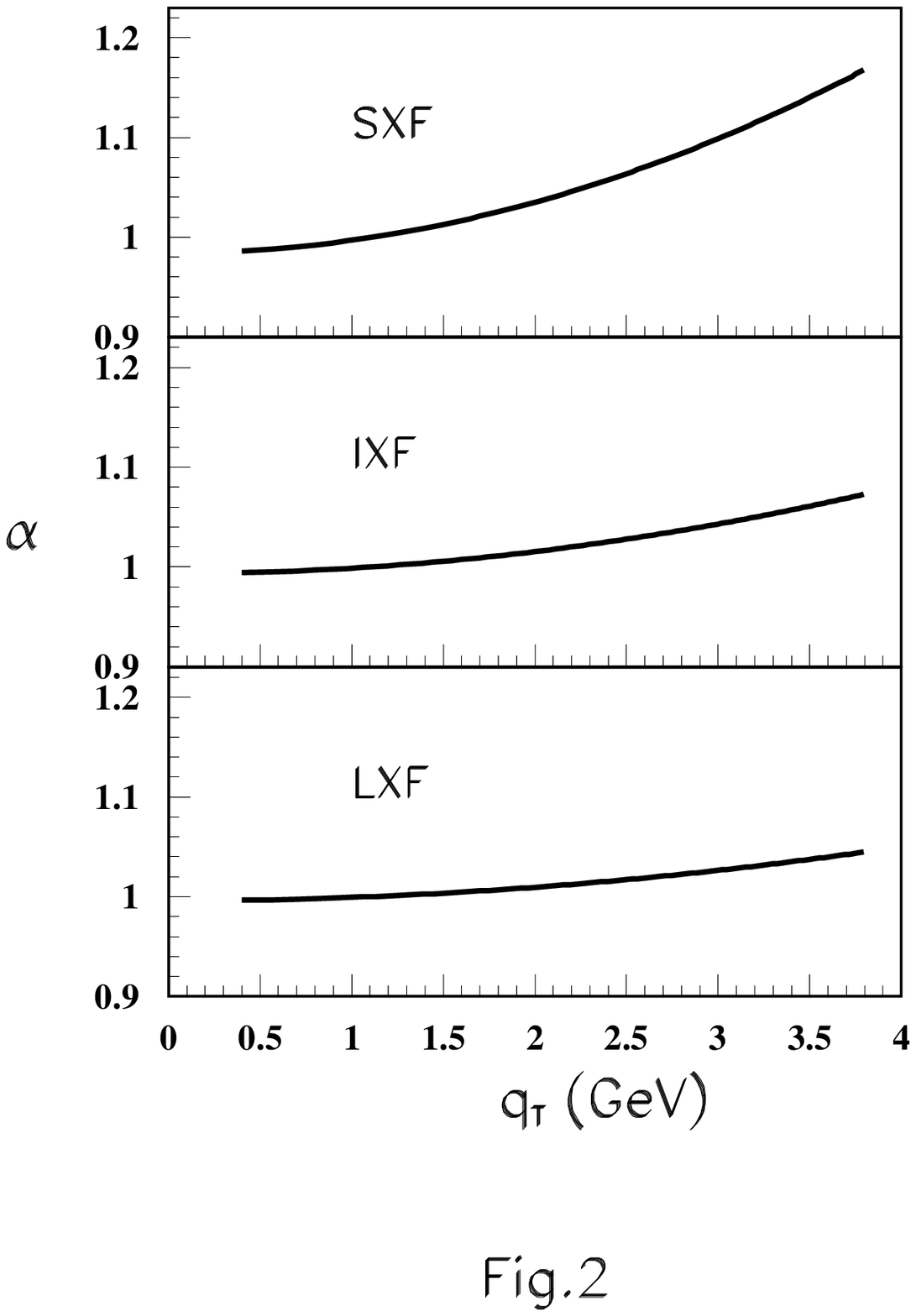,width=1.0in}
\caption{The $\alpha(q_T)$ as a function of $q_T$ for small,
intermediate, and large $x_F$ region, respectively. The curves 
are for $Q=5$ GeV and $800$ GeV beam energy.} 
\label{fig2}
\end{figure}

In Fig.~\ref{fig2}, we plotted our predictions for the value of
$\alpha(q_T)$ defined in Eq.~(\ref{alpha-qt}) as a function of $q_T$
in three different $x_F$ regions.   
From Fig.~\ref{fig2}, it is clear that the shape of $\alpha(q_T)$ in
different $x_F$ ranges are similar to what was observed in recent data
on J/$\Psi$ and $\Psi'$ suppression \cite{JPsi}.  
Such similarity is natural because both $q_T$ spectrum of the Drell-Yan 
and Charmonium production can be approximated by a Gaussian form when
$q_T$ is not too large.  In addition, we emphasize that as shown in
Figs.~\ref{fig1} and \ref{fig2}, even though $\alpha(q_T)$ seems
not to be sensitive to $x_F$, the $x_F$ dependence of transverse
momentum broadening, $\Delta\langle q_T^2\rangle (x_F)$, can be
very strong.

In summary, we derived the $x_F$ dependence of the Drell-Yan
transverse momentum broadening in terms of quark-gluon correlation
functions.  We predicted  that $\Delta \langle q_T^2 \rangle (x_F)$
has a strong $x_F$ dependence. In particular, if the LQS model for the 
correlation functions is valid, $\Delta \langle q_T^2 \rangle
(x_F)$ should have similar $x_F$ dependence as the differential
cross section $d\sigma /dQ^2d x_F$. We also demonstrated that
$\Delta \langle q_T^2 \rangle (x_F)$ is a very sensitive
observable to study the effects of initial state multiple
scattering and the parton energy loss. In fact, $\Delta \langle
q_T^2 \rangle (x_F)$ itself is an excellent observable for
extracting direct information on the nuclear quark-gluon
correlation functions.

\section*{Acknowledgment}

We thank Jianwei Qiu for very helpful discussions. We also thank
M.J. Leitch, J.M. Moss, and J.-C. Peng for useful communications.
W. Zhu is grateful for the hospitality of International Institute
of Theoretical and Applied Physics. This work was supported in
part by the U.S. Department of Energy under Grant Nos.
DE-FG02-87ER40731 and  DE-FG02-96ER40989.

\end{document}